\documentclass[journal=jacsat,manuscript=article]{achemso}

\usepackage[version=3]{mhchem} 
\usepackage{amssymb}

\author{Tae Gwan Park}
\affiliation{Center for Nanophase Materials Sciences, Oak Ridge National Laboratory, Oak Ridge, Tennessee 37831, USA}
\alsoaffiliation{Center for Integrated Nanotechnologies, Los Alamos National Laboratory, Los Alamos, New Mexico 87544, USA}

\author{Xufan Li}
\affiliation{Honda Research Institute USA Inc., San Jose, California 95134, USA}

\author{Kyungnam Kang}
\affiliation{Center for Nanophase Materials Sciences, Oak Ridge National Laboratory, Oak Ridge, Tennessee 37831, USA}

\author{David B. Geohegan}
\affiliation{Department of Materials Science and Engineering, University of Tennessee at Knoxville, Knoxville, Tennessee 37996, USA}

\author{Christopher M. Rouleau}
\affiliation{Center for Nanophase Materials Sciences, Oak Ridge National Laboratory, Oak Ridge, Tennessee 37831, USA}

\author{Alexander A. Puretzky}
\affiliation{Center for Nanophase Materials Sciences, Oak Ridge National Laboratory, Oak Ridge, Tennessee 37831, USA}
\email{puretzkya@ornl.gov}

\author{Kai Xiao}
\affiliation{Center for Nanophase Materials Sciences, Oak Ridge National Laboratory, Oak Ridge, Tennessee 37831, USA}
\email{xiaok@ornl.gov}

\title[An \textsf{achemso} demo]
  {Spontaneous Polarization Suppression of Exciton-Exciton Annihilation in 3R-Stacked MoS$_2$ Bilayers}

\abbreviations{IR,NMR,UV}

\begin{document}

{\footnotesize
\noindent\textbf{Notice:} This manuscript has been authored by UT-Battelle, LLC, under Contract No.\ DE-AC05-00OR22725 with the U.S.\ Department of Energy. The United States Government retains and the publisher, by accepting the article for publication, acknowledges that the United States Government retains a non-exclusive, paid-up, irrevocable, world-wide license to publish or reproduce the published form of this manuscript, or allow others to do so, for United States Government purposes. The Department of Energy will provide public access to these results of federally sponsored research in accordance with the DOE Public Access Plan (\url{http://energy.gov/downloads/doe-public-access-plan}).
}


\newpage

\begin{abstract}
\noindent 
Rapid exciton-exciton annihilation (EEA) in two-dimensional semiconductors limits access to high-density excitonic regimes essential for efficient optoelectronic operation under strong excitation. Here, we show that EEA is suppressed by repulsive dipole-dipole interactions between interlayer excitons polarized by the spontaneous polarization intrinsic to rhombohedral (3R)-stacked MoS$_2$ bilayers. Using ultrafast pump-probe spectroscopy, we measure an EEA rate of $\gamma_{\rm EEA}=(5.03\pm0.99)\times10^{-3}$\,cm$^2$\,s$^{-1}$ in 3R bilayers, which is approximately 18.2-fold smaller than that in monolayers and 2.9-fold smaller than that in nonpolar 2H bilayers. Despite the higher exciton diffusivity recently reported for 3R relative to 2H bilayers, the reduced EEA rate in 3R indicates a rate-limited regime governed by the close-encounter annihilation probability rather than diffusion. A rate-limited annihilation model incorporating a dipole-dipole repulsive potential captures the observed ratio $\gamma_{{\rm EEA},3{\rm R}}/\gamma_{{\rm EEA},2{\rm H}}\approx0.35$ for an exciton-exciton encounter distance of $\sim$1.3 nm, consistent with the bilayer exciton Bohr radius. These results show that spontaneous polarization in 3R-stacked bilayers suppresses nonlinear excitonic losses and provides a route toward high-density excitonics.
\end{abstract}

\section{Introduction}
Reduced dielectric screening and quantum confinement in atomically thin semiconductors markedly enhance Coulomb interactions, stabilizing tightly bound excitons with large binding energies and strong light-matter coupling.\cite{Mak2016NatPhoton,Wang2018RMP} These attributes make two-dimensional (2D) transition metal dichalcogenides (TMDs) a compelling materials platform for photonics and optoelectronics\cite{Mak2016NatPhoton,cheng20212d,autere2018nonlinear,de2025roadmap} -- such as photodetectors, light emitters, and nonlinear optical devices -- enabled by large oscillator strengths and robust excitonic resonances that persist even at room temperature. However, the same low-dimensionality also strengthens many-body interactions among photoexcited quasiparticles. A prominent example is exciton-exciton annihilation (EEA), in which two excitons interact such that one recombines nonradiatively while transferring its energy and momentum to the other, a process that becomes significant at relatively high exciton densities. As excitation densities increase, exciton-exciton encounters become more frequent, activating efficient nonradiative loss pathways that define the upper limit of sustainable exciton density and the efficiency of photonic and optoelectronic devices.\cite{Sun2014NanoLett,Kim2021Science,Vosco2025PRL} In 2D TMDs, EEA is typically much more efficient than in bulk counterparts and conventional 3D semiconductors, introducing a pronounced density-dependent decay channel that rapidly depletes the exciton population and highlights the challenge of sustaining high-density, long-lived excitonic states.\cite{Sun2014NanoLett,Kumar2014PRB,Yuan2015Nanoscale,Yu2016PRB,Goodman2020JPCC,Chen2022NanoRes,Li2024AdvMater,Vosco2025PRL}

Mitigating EEA is therefore crucial not only for efficient optoelectronic operation across a wide range of excitation densities\cite{Kim2021Science,Uddin2022NanoLett_EL,Uddin2022ACSNano_Bilayer} but also for accessing strongly correlated excitonic phases in 2D semiconductors.\cite{Arp2019NatPhoton,Yu2019ACSNano_EHL,Yu2023ACSNano_PhaseDiagram,Sortino2023LSA,Xu2025PRL_ASE} Reported strategies include detuning band-structure resonances through strain,\cite{Kim2021Science,Uddin2022NanoLett_EL,Uddin2022ACSNano_Bilayer} modifying Coulomb screening via dielectric engineering,\cite{Hoshi2017PRB,Steinhoff2021PRB} and exploiting metasurfaces,\cite{Sortino2023LSA,Yuan2021NanoLett_Metasurface} and van der Waals heterostructures.\cite{Cai2024NanoLett_Moire} In particular, spatial separation of carriers through formation of dipolar excitons, e.g., interlayer excitons with out-of-plane dipole moment in bilayer semiconductors, can alter excitonic transport and many-body dynamics through dipole-dipole repulsion.\cite{Unuchek2019NatNano,Li2020NatMater_Dipolar,Sun2022NatPhoton_Transport,Yuan2023ACSNano_1D,Agunbiade2025ACSNano_Superdiffusive} Such repulsive interactions could, in principle, introduce an effective interaction barrier that suppresses the short-range exciton-exciton encounters required for annihilation, but their quantitative impact on EEA remains largely unexplored.

Rhombohedral (3R) MoS$_2$ bilayers intrinsically host dipolar excitons via spontaneous out-of-plane polarization.\cite{Yang2022NatPhoton_PV,Liang2022PRX_Asymmetric,Dong2023NatNano_Piezophoto} In 3R stacking, broken inversion symmetry produces an intrinsic interlayer potential that layer-polarizes K valley electronic states, reduces electron-hole overlap, and yields an out-of-plane exciton dipole moment,\cite{Agunbiade2025ACSNano_Superdiffusive,Yang2022NatPhoton_PV,Liang2022PRX_Asymmetric,Dong2023NatNano_Piezophoto,Agunbiade2024PRB_TA} offering a route to investigate the role of dipolar repulsion in EEA. Despite extensive studies of exciton dynamics in 3R MoS$_2$ bilayers,\cite{Chen2022NanoRes,Agunbiade2025ACSNano_Superdiffusive,Agunbiade2024PRB_TA,Zhou2025ACSNano_TwistSynthesis,Zhong2025NanoLett_Mobility} isolating and quantifying the contribution of dipolar repulsion to EEA remains experimentally challenging. First, experimentally resolving EEA requires an intermediate excitation density -- above the low densities where bimolecular annihilation is negligible\cite{Agunbiade2024PRB_TA,Zhong2025NanoLett_Mobility} but below sufficiently high densities at which competing ultrafast channels, such as defect-assisted or Auger recombination processes, obscure EEA in time-domain measurements.\cite{Zhou2025ACSNano_TwistSynthesis} Second, quantitative attribution to dipolar repulsion cannot rely on monolayer-bilayer comparisons, because bilayer TMDs introduce additional layer-number effects, most notably indirect-gap relaxation channels\cite{Zhou2025ACSNano_TwistSynthesis} and associated nonradiative pathways,\cite{Yuan2015Nanoscale,Chen2022NanoRes} that can modify exciton lifetimes. A more appropriate control is the centrosymmetric 2H bilayer, which shares the same chemical composition and thickness but lacks built-in polarization. Third, EEA is highly sensitive to extrinsic factors such as dielectric screening,\cite{Yu2016PRB,Goodman2020JPCC} local strain,\cite{Kim2021Science,Uddin2022NanoLett_EL,Uddin2022ACSNano_Bilayer} defect density,\cite{Liu2019Nanoscale_WS2Defect,Soni2025SciRep_Substrate,Zhang2021ACSPhotonics_DefectEEA} necessitating a side-by-side comparison of 2H and 3R bilayers with closely matched crystalline quality and dielectric environment. Finally, while encapsulation with h-BN is often employed to improve optical properties, it can itself suppress EEA through enhanced dielectric screening and reduced disorder,\cite{Hoshi2017PRB,Steinhoff2021PRB} thereby making EEA difficult to identify even at relatively high excitation densities.\cite{Agunbiade2025ACSNano_Superdiffusive} These constraints have so far hindered an investigation of how spontaneous polarization and the resulting dipolar repulsion modulates the EEA rate in 3R-stacked bilayers.

In this work, we directly isolate the role of spontaneous polarization in EEA by comparing 2H and 3R MoS$_2$ bilayers grown on the same substrate that are confirmed to have comparable quality. By carefully probing fluence-dependent transient reflectance within an intermediate pump-fluence range, we directly compare EEA rates in 2H and 3R MoS$_2$ bilayers and find that the EEA rate in 3R bilayers is $\sim$2.9 fold smaller than that in 2H. Although 3R bilayers exhibit higher exciton diffusivity than 2H bilayers,\cite{Agunbiade2025ACSNano_Superdiffusive} the reduced EEA in 3R bilayers indicates a rate-limited regime where the close-encounter annihilation probability, rather than diffusion, controls the EEA rate. A rate-limited annihilation model incorporating short-range repulsive dipole-dipole interactions between layer-polarized excitons quantitatively accounts for the observed suppression, underscoring the role of spontaneous polarization in governing exciton-exciton interactions. These results show that 3R stacking and the associated built-in polarization provide an intrinsic route to suppress nonlinear excitonic losses and expand the accessible high-density and long-lived exciton regime in 2D semiconductors.

\section{Results and discussion}
\subsection{Structural and electronic properties of 2H and 3R bilayer MoS$_2$}

\begin{figure}[!b]
  \centering
  \includegraphics[width=0.5\linewidth]{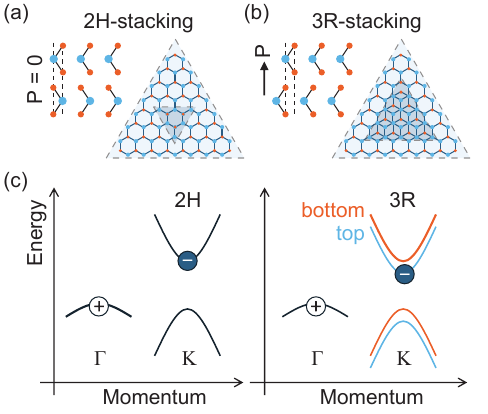}
  \caption{Stacking-dependent symmetry and electronic structure of MoS$_2$ bilayers.
(a) 2H (AA$'$) stacking with 180$^\circ$ rotation between layers, restoring inversion symmetry and yielding zero out-of-plane polarization ($P = 0$). (b) 3R (AB) stacking with a lateral shift of one-third of the unit cell, breaking inversion symmetry and inducing an out-of-plane polarization ($P \neq 0$; opposite for BA stacking). (c) Schematic band structure of 2H and 3R bilayers.}
  \label{fig1}
\end{figure}

Both hexagonal 2H and rhombohedral 3R bilayers occur in natural deposits.\cite{He2014PRB_Stacking,Paradisanos2020NatCommun} As illustrated in Fig.~\ref{fig1}(a), the 2H bilayer comprises two layers rotated by 180$^\circ$ relative to each other, which restores inversion symmetry for an even number of layers. In contrast, in 3R bilayer [Fig.~\ref{fig1}(b)], adjacent layers share the same crystallographic orientation but are laterally shifted by one-third of the unit cell (i.e., the metal sublattice in one layer aligns above the chalcogen sublattice in another layer). This breaks inversion symmetry and gives rise to an out-of-plane spontaneous polarization\cite{Yang2022NatPhoton_PV,Liang2022PRX_Asymmetric,Dong2023NatNano_Piezophoto}. The lack of inversion symmetry in 3R bilayers enables nonlinear optical response,\cite{Shi2017AdvMater_NLO,Zhang2020AdvMater_SHG} ferroelectric behavior,\cite{Dong2023NatNano_Piezophoto,Li2024NatCommun_Memory,Yang2024NatCommun_Switching} ultrafast spontaneous photovoltaic effect,\cite{Wu2022SciAdv_UltrafastPV} and nanosecond ferroelectric switching of intralayer excitons.\cite{Liang2025PRX_Switching} These symmetry differences also produce distinct electronic band structures as shown in Fig.~\ref{fig1}(c). In 3R bilayers, the built-in polarization generates an intrinsic interlayer potential and asymmetric interlayer coupling that lifts the layer degeneracy at the K valleys, effectively resulting in a type-II alignment.\cite{Yang2022NatPhoton_PV,Liang2022PRX_Asymmetric} By contrast, in 2H bilayers, the K valley conduction and valence bands remain layer-degenerate and show no net layer selectivity in the electronic eigenstates.\cite{Agunbiade2024PRB_TA,Agunbiade2025ACSNano_Superdiffusive}

Because bilayer TMDs are indirect-gap semiconductors, the valence-band maximum shifts toward the $\Gamma$ valley. In 3R MoS$_2$ bilayers, photoexcited electrons in the K valley become layer-selective and preferentially occupy the lower-energy K-point conduction-band edge of one layer (top layer, for AB stacking), while holes accumulate near the $\Gamma$-valley valence-band edge [Fig.~\ref{fig1}(c)].\cite{Agunbiade2024PRB_TA} Meanwhile, the valence states at $\Gamma$ remain strongly hybridized across layers, which reduces hole layer polarization relative to electrons. The resulting layer-selective electronic structure enhances electron localization to a specific layer in 3R, reduces electron-hole wavefunction overlap, and imparts dipolar exciton character. Consistent with this picture, prior ultrafast spectroscopy studies have reported prolonged exciton lifetimes\cite{Agunbiade2024PRB_TA} and superdiffusive exciton transport\cite{Agunbiade2025ACSNano_Superdiffusive} in 3R-stacked MoS$_2$ bilayers.

\subsection{Sample preparation and optical characterization}

\begin{figure}[!b]
  \centering
  \includegraphics[width=0.5\linewidth]{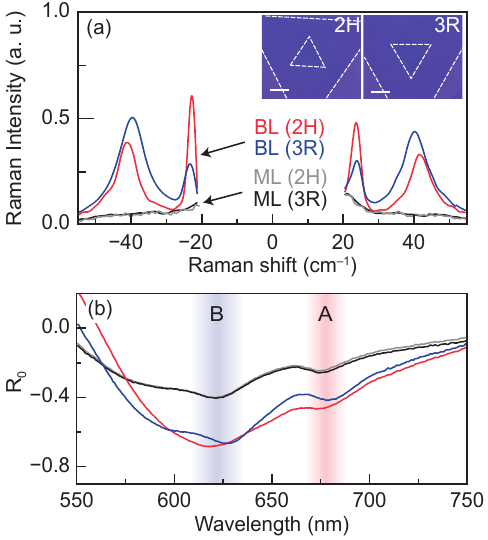}
  \caption{Optical identification of stacking configuration in MoS$_2$ bilayers. (a) Low-frequency Raman spectra of monolayer (ML) and bilayer (BL) 2H and 3R MoS$_2$, labeled ML (2H), ML (3R), BL (2H), and BL (3R). Two interlayer vibrational modes are observed in the BL samples: the in-plane shear mode at $\sim$24\,cm$^{-1}$ and the out-of-plane breathing mode at $\sim$40\,cm$^{-1}$. Insets: Optical microscope images of 2H and 3R samples. The inner white dashed triangle outlines the BL region, while the area up to the outer white dashed boundary corresponds to the ML region. Scale bars: 10 $\mu$m. (b) Corresponding normalized reflectance spectra. Shaded regions indicate the A and B excitonic resonances.}
  \label{fig2}
\end{figure}

Bilayer MoS$_2$ crystals were grown on SiO$_2$/Si substrates by chemical vapor deposition (CVD) as described in the Methods section. Because the thermodynamically stable bilayer stacking configurations of MoS$_2$ are 2H and 3R,\cite{He2014PRB_Stacking,Paradisanos2020NatCommun} the as-grown samples naturally contain both 2H- and 3R-stacked bilayer domains on a single substrate. The stacking configuration was identified from optical microscope images of as-grown bilayer domains, based on the relative rotation angles formed during CVD, as shown in the inset of Fig.~\ref{fig2}(a). The typical edge length of the top layer in the bilayer region is $\sim$20 $\mu$m. Importantly, the coexistence of large-area, uniform 2H and 3R bilayers on one substrate enables direct side-by-side optical comparisons under identical experimental conditions.

Optical characterization was performed using Raman spectroscopy and optical reflectance at room temperature under ambient conditions. Raman spectra were acquired using a 532 nm continuous-wave laser (beam spot $\sim$1 $\mu$m), and reflectance spectra were measured using a white-light continuum source (beam spot $\sim$5 $\mu$m) (see Methods). Low-frequency Raman spectroscopy further confirms the stacking configuration [Fig.~\ref{fig2}(a)]. In the low-frequency range ($< 60$\,cm$^{-1}$), two pronounced peaks appear only in bilayer MoS$_2$ and are assigned to the interlayer shear (SM) and breathing (BM) modes, consistent with previous observations.\cite{Huang2016NanoLett_Raman,Yan2015NanoLett_Trilayer} The SM lies at $\sim$24\,cm$^{-1}$ for both 2H and 3R bilayers, while the BM peak occurs at higher frequency in 2H bilayers ($\sim$42\,cm$^{-1}$) than in 3R bilayers ($\sim$40\,cm$^{-1}$). The softening of the BM in 3R indicates a slightly weaker interlayer restoring force compared to 2H, consistent with stronger interlayer coupling in 2H.\cite{Huang2016NanoLett_Raman,Yan2015NanoLett_Trilayer} Full Raman spectra are provided in Fig.~S1 in the Supporting Information.

Figure~\ref{fig2}(b) shows the normalized reflectance spectrum, $R_0 = R_{\mathrm{sample}}/R_{\mathrm{substrate}}$, from monolayers (ML) and 2H- and 3R-stacked bilayers, where the excitonic resonances appear as pronounced spectral features. For MoS$_2$ on SiO$_2$/Si substrate, $R_0$ typically shows dips near the exciton resonances due to absorption- and phase-induced changes in MoS$_2$ combined with interference effects in the SiO$_2$ layer.\cite{Niu2018Nanomaterials} Across the monolayer, 2H, and 3R bilayer samples, two resonances near $\sim$680 nm and 625 nm correspond to the A and B excitons, respectively. Spectra acquired from monolayer regions adjacent to the 2H and 3R bilayer domains are nearly identical, indicating similar crystal quality. Compared with ML, bilayers display stronger and spectrally modified excitonic responses. Subtle differences between 2H and 3R bilayers are also observed, reflecting stacking-dependent interlayer coupling and electronic structure.\cite{Paradisanos2020NatCommun}

\subsection{Excitation density-dependent ultrafast pump-probe spectroscopy}

\begin{figure}[!b]
  \centering
  \includegraphics[width=1.0\linewidth]{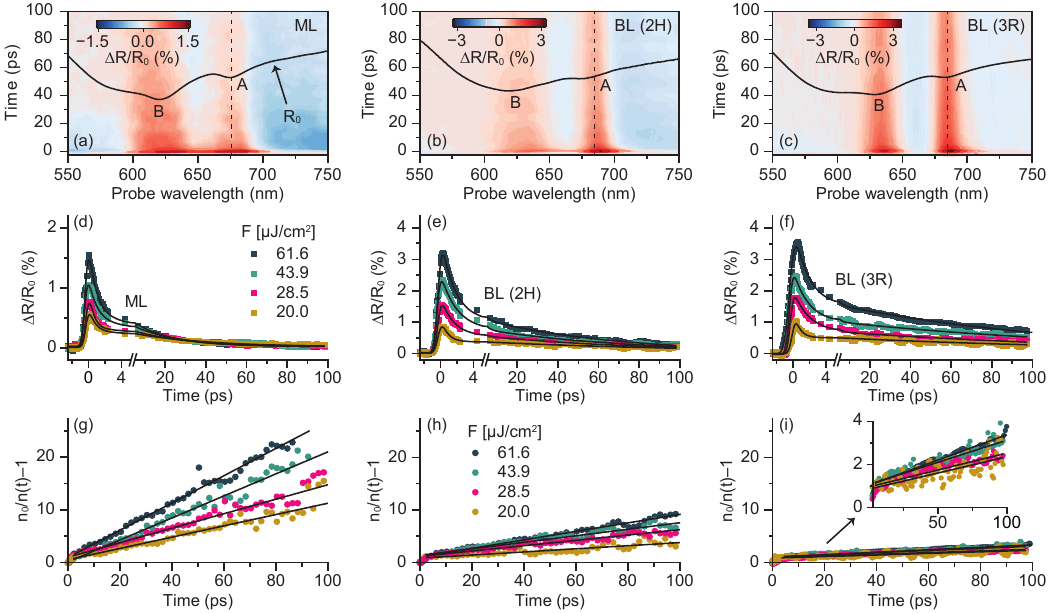}
  \caption{Ultrafast exciton dynamics of monolayer and bilayer MoS$_2$ at varying pump fluence. (a--c) Transient reflectance maps, $\Delta R/R_0$, as a function of probe wavelength and pump-probe time delay for (a) monolayer, (b) 2H bilayer, and (c) 3R bilayer measured at a pump fluence of 61.6 $\mu$J\,cm$^{-2}$. The black curve shows the linear reflectance spectrum $R_0$ in Fig.~\ref{fig2}(b) with arbitrary scaling. The vertical dashed lines indicate the A-exciton resonance. (d--f) Corresponding $\Delta R/R_0$ kinetics extracted at the A-exciton bleach wavelength at different pump fluences, with bi-exponential fits (black curves). (g--i) Corresponding time evolution of $n_0/n(t)-1$. Solid black lines are linear fits in the EEA-dominated regime ($t \gtrsim 3$ ps), with slopes $n_0 \gamma_{\mathrm{EEA}}$. The inset in (i) enlarges the 3R data for clarity.}
  \label{fig3}
\end{figure}

Figure~\ref{fig3} summarizes the stacking-configuration- and excitation-density-dependent exciton kinetics in monolayer, 2H-, and 3R-stacked MoS$_2$ bilayers measured by ultrafast pump-probe spectroscopy at room temperature under ambient conditions. Using a 400 nm pump and a broadband white-light continuum probe, we measured the transient reflectance, $\Delta R/R_0$, as a function of probe wavelength and pump-probe time delay ($t$), as shown in Figs.~\ref{fig3}(a--c). Both pump and probe spot sizes were $\sim$5 $\mu$m at the sample, and the pump fluence ($F_{\mathrm{pump}}$) was varied from $\sim$20 to 61\,$\mu$J\,cm$^{-2}$ using a set of neutral-density optical filters (see Methods for details). Over this $F_{\mathrm{pump}}$ range, the initial exciton density per layer is estimated as $n_0\,=\,(1.0-3.1)\,\times\,10^{12}$\,cm$^{-2}$ by accounting for optical absorption of MoS$_2$ and substrate interference effects (see Methods) and assuming one absorbed photon creates one electron-hole pair.

For the monolayer in Fig.~\ref{fig3}(a), the signal was acquired from a monolayer region adjacent to the 2H bilayer domain. We confirmed that its pump-probe response is nearly identical to that of a monolayer region adjacent to the 3R domain (Fig.~S2). The black curves overlaid on the transient maps in Figs.~\ref{fig3}(a--c) are the linear reflectance spectra, $R_0$ in Fig.~\ref{fig2}(b), plotted with arbitrary scaling as a visual guide to the excitonic resonance positions. As indicated by the vertical dashed lines, the positive $\Delta R/R_0$ signal near 680 nm is assigned to photoinduced bleaching of the K valley A-exciton resonance, and therefore it tracks the kinetics of the K valley carrier population, whose amplitude is approximately proportional to the exciton density.\cite{Agunbiade2025ACSNano_Superdiffusive,Agunbiade2024PRB_TA} Figures~\ref{fig3}(d--f) show the excitation-density-dependent kinetics extracted at this probe wavelength. Across all samples, the peak $\Delta R/R_0$ scales approximately linearly with $F_{\mathrm{pump}}$ (Fig.~S3).

For each sample, the transients are well described by a bi-exponential decay. The fast and slow time constants ($\tau_1$ and $\tau_2$) extracted at each $F_{\mathrm{pump}}$ are summarized in Table~S1 in the Supporting Information. In all samples, $\tau_1 \approx 1.5$ ps, which we attribute primarily to rapid hot-exciton cooling and fast exciton capture by defects.\cite{Nie2014ACSNano_Thermalization,Wang2015NanoLett_Defect,Das2021LSA_Subgap} The slower component, $\tau_2$, which dominates for $t \gtrsim 3$ ps after the $\tau_1$ process, is strongly modulated by both stacking and excitation density. In the monolayer, $\tau_2$ decreases from $\sim$37 ps at low excitation density ($F_{\mathrm{pump}} = 20$ $\mu$J\,cm$^{-2}$) to $\sim$19 ps at high excitation density ($F_{\mathrm{pump}} = 61.6$ $\mu$J\,cm$^{-2}$). In the 2H bilayer, $\tau_2$ decreases from $\sim$95 ps to $\sim$64 ps over the same fluence range. In the 3R bilayer, $\tau_2$ shows a more modest reduction, decreasing from $\sim$160 ps to $\sim$135 ps as $F_{\mathrm{pump}}$ increases from 20 to 61.6 $\mu$J\,cm$^{-2}$. Such reductions in $\tau_2$ with increasing excitation density are consistent with density-dependent nonradiative Auger processes, i.e., EEA.\cite{Sun2014NanoLett,Yuan2015Nanoscale,Yu2016PRB,Chen2022NanoRes} Notably, $\tau_2$ also depends strongly on stacking: at a fixed $F_{\mathrm{pump}}$, $\tau_2$ in 3R is nearly twice that in 2H. This longer lifetime in 3R is consistent with prior reports,\cite{Agunbiade2024PRB_TA} which attributed the longer recombination lifetime to reduced electron-hole wavefunction overlap arising from the built-in polarization in 3R bilayers. The $F_{\mathrm{pump}}$-dependent kinetics across monolayer, 2H, and 3R enable a quantitative extraction of the EEA rate while disentangling layer-number and stacking effects, as discussed next.

\subsection{Exciton-exciton annihilation rate}

In the presence of EEA, the exciton population kinetics $n(t)$ can be described by a rate equation for bimolecular recombination,

\begin{equation}
\frac{dn}{dt}=-\gamma_1 n-\gamma_{\mathrm{EEA}} n^2,
\label{eq:rate}
\end{equation}

where $n$ is the exciton density (cm$^{-2}$), $\gamma_1$ is the first-order recombination rate constant (s$^{-1}$), and $\gamma_{\mathrm{EEA}}$ is the EEA rate (cm$^2$\,s$^{-1}$). At early times and sufficiently high densities such that $\gamma_{\mathrm{EEA}} n \gg \gamma_1$ (typically within a few to 100 ps),\cite{Sun2014NanoLett,Kumar2014PRB,Yu2016PRB,Chen2022NanoRes,Soni2025SciRep_Substrate,Zhang2021ACSPhotonics_DefectEEA} the EEA term dominates, and the solution reduces to
\begin{equation}
\frac{n_0}{n(t)}-1=\gamma_{\mathrm{EEA}} n_0 t,
\label{eq:eea_linear}
\end{equation}

a form widely used to analyze EEA-driven decay dynamics.\cite{Sun2014NanoLett,Kumar2014PRB,Yuan2015Nanoscale,Yu2016PRB} Figures~\ref{fig3}(g--i) show the exciton kinetics in terms of $n_0/n(t)-1$. The data exhibit linear behavior for $t \gtrsim 3$ ps, with slopes of $\gamma_{\mathrm{EEA}} n_0$, consistent with an EEA-dominated regime in this window. The extracted $\gamma_{\mathrm{EEA}}$ values were insensitive, within fitting uncertainty, to reasonable variations of the fit start time in the 3--5 ps range. Using Eq.~\ref{eq:eea_linear} and $n_0$, EEA rates are estimated as $\gamma_{\mathrm{EEA}} = (9.14 \pm 0.85) \times 10^{-2}$ cm$^2$\,s$^{-1}$ for monolayer (ML), $(1.43 \pm 0.37) \times 10^{-2}$ cm$^2$\,s$^{-1}$ for the 2H bilayer, and $(5.03 \pm 0.99) \times 10^{-3}$\,cm$^2$\,s$^{-1}$ for the 3R bilayer (Table~S2). While the absolute $\gamma_{\mathrm{EEA}}$ values depend on the estimated $n_0$, the observed suppression of $\gamma_{\mathrm{EEA}}$ in 3R relative to 2H is more robust because both bilayer domains were measured on the same substrate under identical optical conditions. The monolayer value is comparable to previously reported ranges of $4 \times 10^{-2}$--$7 \times\,10^{-2}$\,cm$^2$ s$^{-1}$ for unencapsulated MoS$_2$ monolayers.\cite{Sun2014NanoLett,Yu2016PRB,Li2024AdvMater,Tsai2020ACSOmega} Differences among reported $\gamma_{\mathrm{EEA,ML}}$ values (including ours) are commonly attributed to dielectric screening,\cite{Yu2016PRB,Goodman2020JPCC} local strain,\cite{Kim2021Science,Uddin2022NanoLett_EL,Uddin2022ACSNano_Bilayer} and defect density.\cite{Soni2025SciRep_Substrate,Zhang2021ACSPhotonics_DefectEEA,Liu2019Nanoscale_WS2Defect} These extrinsic factors are unlikely to dominate the present bilayer comparison because the 2H and 3R domains were grown on the same substrate and exhibit nearly identical optical characteristics; accordingly, adjacent monolayer regions show nearly identical EEA behavior (Fig.~S2), and we therefore do not invoke extrinsic differences to explain the bilayer trend.

Both bilayers exhibit markedly smaller slopes than the monolayer [Figs.~\ref{fig3}(g--i)] yielding reduced EEA rates: $\gamma_{\mathrm{EEA},2\mathrm{H}} = (1.43 \pm 0.37) \times 10^{-2}$\,cm$^2$\,s$^{-1}$ and $\gamma_{\mathrm{EEA},3\mathrm{R}} = (5.03 \pm 0.99) \times 10^{-3}$\,cm$^2$\,s$^{-1}$. Compared to the monolayer, $\gamma_{\mathrm{EEA}}$ is reduced by a factor of $\sim$6.4 in 2H and $\sim$18.2 in 3R. Suppressed EEA in bilayer TMDs compared with monolayers is well established,\cite{Yuan2015Nanoscale,Chen2022NanoRes,Li2024AdvMater} and is generally attributed to two effects: (1) enhanced dielectric screening in thicker layers, which weakens Coulomb-mediated many-body interactions, and (2) the emergence of an indirect gap in bilayers, which opens a phonon-assisted (momentum-conserving) annihilation pathway that is intrinsically less probable than direct EEA in monolayers.

These mechanisms account for the overall EEA reduction from monolayer to bilayer but do not by themselves explain the additional stacking dependence between 2H and 3R bilayers. First, screened Coulomb interactions in 2D semiconductors are governed by the dielectric environment and the 2D polarizability $\chi_{\mathrm{2D}}$.\cite{Cudazzo2011PRB_Screening} In our experiments, the environment is identical for all regions (same substrate), and $\chi_{\mathrm{2D}}$ in MoS$_2$ scales approximately linearly with layer number, i.e., $\chi_{\mathrm{2D}}(\mathrm{bilayer}) \approx 2\chi_{\mathrm{2D}}(\mathrm{monolayer})$, for both 2H and 3R bilayers.\cite{Tian2019NanoLett_Polarizability,Ferreira2022PRB_Susceptibility} Thus, the enhancement of screening relative to the monolayer should be comparable in 2H and 3R and cannot explain their different $\gamma_{\mathrm{EEA}}$. This is also consistent with microscopic theory indicating that EEA relies strongly on large-momentum Coulomb scattering where environmental screening plays a weaker role than intrinsic TMD polarization.\cite{Steinhoff2021PRB}

Second, the phonon-assisted indirect channel should depend primarily on the indirect-gap energy and the relevant phonon phase space. In MoS$_2$, 2H and 3R bilayers have similar indirect-gap energies,\cite{Agunbiade2024PRB_TA,Liu2014NatCommun_TwistCoupling} and their phonon dispersions and density of states are essentially the same over a broad frequency range.\cite{Huang2016NanoLett_Raman,Coutinho2017JPCS} The main stacking-dependent phonon differences appear in the low-frequency interlayer shear and breathing modes near the zone center [Fig.~\ref{fig2}(a)],\cite{Huang2016NanoLett_Raman,Yan2015NanoLett_Trilayer} whereas indirect (intervalley) processes typically require large in-plane momentum transfer and are dominated by finite-momentum intralayer phonons near the zone edge, not zone-center interlayer modes.\cite{Hu2023AdvFunctMater_UED} Consequently, phonon-mediated contributions to indirect EEA should be comparable in 2H and 3R bilayers. This interpretation is consistent with prior ultrafast studies in which EEA-related exciton dynamics are effectively absent from the measured time window.\cite{Zhou2025ACSNano_TwistSynthesis,Zhong2025NanoLett_Mobility} For instance, transient absorption measurements on twisted bilayer MoS$_2$ performed at pump fluences $\sim$30--100 times higher than ours report that the early-time decay (on the order of a few ps) is dominated by ultrafast nonradiative channels, often attributed to defect-assisted and/or Auger-like processes, making it difficult to isolate EEA-related kinetics explicitly.\cite{Zhou2025ACSNano_TwistSynthesis} In that high-density regime, the subsequently observed slower dynamics are instead discussed in terms of phonon-assisted exciton relaxation and recombination that depend primarily on interlayer coupling (and thus interlayer distance), which can lead to comparable exciton lifetimes for 2H and 3R stackings.\cite{Zhou2025ACSNano_TwistSynthesis} Moreover, ultrafast spectroscopy measurements on twisted WS$_2$ bilayers performed at pump fluences approximately 180 times lower than those used in our study, where EEA is not expected to contribute significantly, have also reported comparable exciton lifetimes for the 2H and 3R configurations.\cite{Zhong2025NanoLett_Mobility}

In contrast, pump-probe measurements performed in a comparatively weak excitation regime (up to $\sim$30\,$\mu$J\,cm$^{-2}$, i.e., roughly half of our pump fluence) reported that the 3R bilayer exhibits an exciton lifetime approximately twice as long as that of the 2H bilayer, which was attributed to reduced electron-hole wavefunction overlap in the 3R stacking.\cite{Agunbiade2024PRB_TA} In that low-density regime, the exciton decay shows little fluence dependence, consistent with the expectation that EEA does not contribute significantly, similar to our observations, where the exciton dynamic traces at 20 and 28\,$\mu$J\,cm$^{-2}$ do not differ markedly  (Fig.~\ref{fig3}). A similar absence of systematic pump-fluence dependence has also been reported even at higher excitation densities (up to pump fluence $\sim$60\,$\mu$J\,cm$^{-2}$, comparable to our conditions), where fluence-dependent lifetime trends are difficult to resolve.\cite{Agunbiade2025ACSNano_Superdiffusive} We attribute this apparent insensitivity, at least in part, to EEA suppression induced by h-BN encapsulation, which can enhance dielectric screening and reduce disorder, thereby hindering the identification of EEA contributions even at relatively high excitation densities.\cite{Hoshi2017PRB,Steinhoff2021PRB}

Our intermediate excitation density, combined with the absence of h-BN encapsulation, places the present measurements in a regime where EEA-related exciton dynamics can be resolved within the relevant time window. Although EEA-independent channels, including radiative recombination and phonon-assisted relaxation, also contribute to the transient decay, they are expected to show only weak pump-fluence dependence over our experimental range and to remain broadly similar between the 2H and 3R bilayers.\cite{Zhou2025ACSNano_TwistSynthesis,Zhong2025NanoLett_Mobility,Liu2014NatCommun_TwistCoupling} The pump-fluence-dependent component of the dynamics therefore primarily reflects EEA, allowing a direct comparison of the EEA rates between the two stackings. The further suppression of EEA observed in the 3R bilayer thus indicates that stacking configuration provides an additional degree of freedom for modulating EEA beyond layer-number effects. A key feature unique to 3R MoS$_2$ is its broken inversion symmetry, which generates out-of-plane spontaneous polarization and promotes spatial separation of electrons and holes. The resulting dipolar excitons can significantly influence nonequilibrium exciton dynamics, consistent with previous reports of prolonged exciton lifetimes\cite{Agunbiade2024PRB_TA} and superdiffusive exciton transport\cite{Agunbiade2025ACSNano_Superdiffusive} in 3R MoS$_2$ bilayers. In our case, such aligned dipole moments can give rise to repulsive dipole-dipole interactions, thereby reducing the probability of the close exciton-exciton encounters required for Auger-type annihilation.\cite{Knorr2002PRB_Adsorbate,Yokoyama2007PRL_DipolarMolecules} Below, we develop a microscopic picture for the reduced EEA in 3R MoS$_2$ based on these repulsive dipolar interactions.

\subsection{Dipolar repulsion and EEA suppression in 3R MoS$_2$}

The effective bimolecular EEA rate $\gamma_{\mathrm{EEA}}$ can be expressed in terms of a diffusion-limited contribution $k_D$ and a rate-limited contribution $k_R$ as\cite{Yuan2015Nanoscale}

\begin{equation}
\frac{1}{\gamma_{\mathrm{EEA}}}=\frac{1}{k_D}+\frac{1}{k_R}.
\label{eq:kdkr}
\end{equation}

Here, $k_D$ characterizes mutual diffusion of excitons ($k_D \propto D$, where $D$ is the exciton diffusion coefficient),\cite{Yu2016PRB,Goodman2020JPCC} whereas $k_R$ captures the intrinsic (rate-limited) annihilation rate once two excitons are sufficiently close. For monolayer and multilayer TMDs, overall exciton diffusion is an order of magnitude faster than annihilation,\cite{Yuan2015Nanoscale} so that $\gamma_{\mathrm{EEA}}$ is expected to be rate-limited ($\gamma_{\mathrm{EEA}} \approx k_R$). Our stacking-dependent data further support this view: recent measurements show that $D$ in 3R can be an order of magnitude larger than in 2H MoS$_2$,\cite{Agunbiade2025ACSNano_Superdiffusive} which, under a purely diffusion-limited picture ($\gamma_{\mathrm{EEA}} \approx k_D \propto D$), would predict a larger $\gamma_{\mathrm{EEA}}$ in 3R. Instead, we observe the opposite trend (suppressed EEA in 3R relative to 2H), indicating that the intrinsic annihilation probability, rather than diffusive transport, governs the effective EEA rates in our measurements, which is consistent with prior reports of only a weak dependence of $\gamma_{\mathrm{EEA}}$ on exciton diffusion.\cite{Yuan2015Nanoscale,Goodman2020JPCC}

\begin{figure}[!b]
  \centering
  \includegraphics[width=0.5\linewidth]{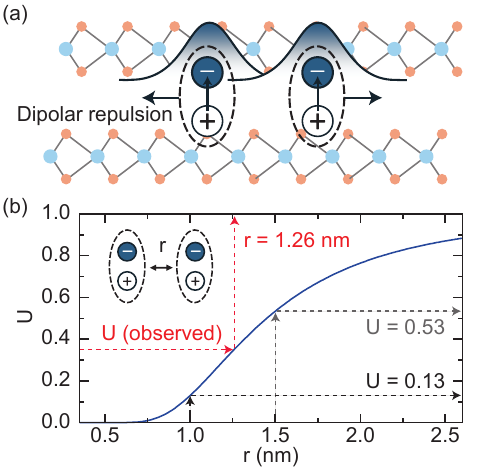}
  \caption{Schematic of EEA suppression in 3R MoS$_2$ bilayers. (a) Illustration of dipole-dipole repulsion, where spontaneous polarization promotes layer-polarized, dipolar excitons. (b) Calculated Boltzmann suppression factor, $U = \exp[-V_{\mathrm{dd}}(r)/k_B T]$, as a function of the in-plane exciton-exciton separation $r$. The red dashed line indicates the measured EEA suppression in 3R relative to the nonpolar 2H bilayer and the corresponding separation ($r \approx 1.26$ nm). The black and gray dashed lines mark plausible separations ($r = 1.0$ and 1.5 nm) and their corresponding suppression factors ($U = 0.13$ and 0.53), respectively.}
  \label{fig4}
\end{figure}

Within the rate-limited picture, the repulsive dipolar interactions in 3R MoS$_2$ reduce close exciton-exciton encounters as illustrated in Fig.~\ref{fig4}(a) and thus further suppress EEA compared to 2H MoS$_2$, where such dipolar repulsion is absent. For two-body interactions, the exciton-exciton encounter probability distribution, $f$, in the presence of the dipole-dipole potential, $V_{\mathrm{dd}}(r)$, can be written as\cite{Knorr2002PRB_Adsorbate,Yokoyama2007PRL_DipolarMolecules}

\begin{equation}
f = U f_0,
\label{eq:fboltz}
\end{equation}

where $r$ is the exciton-exciton separation, $f_0$ is the probability distribution in the absence of dipolar repulsion, and $U = e^{-V_{\mathrm{dd}}(r)/k_B T}$ is the Boltzmann factor associated with dipolar repulsion, where $k_B$ is the Boltzmann constant and $T = 300$ K. Here, we take the 2H bilayer as the nonpolar reference, for which the short-range encounter probability is represented by $f_0$. Assuming the effective EEA rate scales with the close-encounter probability in the rate-limited regime, the EEA rate in 3R relative to 2H is expected to be suppressed by a factor given by

\begin{equation}
\frac{\gamma_{\mathrm{EEA,3R}}}{\gamma_{\mathrm{EEA,2H}}} \approx \frac{f}{f_0} = U.
\label{eq:ratioU}
\end{equation}

To model the repulsive dipolar interactions $V_{\mathrm{dd}}(r)$, we consider each exciton as an out-of-plane electric dipole arising from the average vertical electron-hole separation ($d$), giving

\begin{equation}
V_{\mathrm{dd}}(r)=\frac{e^2}{2\pi\epsilon_0\kappa}\left(\frac{1}{r}-\frac{1}{\sqrt{r^2+d^2}}\right),
\label{eq:Vdd}
\end{equation}

where $\kappa = (\varepsilon_{\mathrm{top}} + \varepsilon_{\mathrm{bottom}})/2$ is the environmental dielectric constant. For our geometry, $\varepsilon_{\mathrm{top}} = \varepsilon_{\mathrm{air}} = 1$ and $\varepsilon_{\mathrm{bottom}} = \varepsilon_{\mathrm{SiO_2}} = 3.9$,\cite{Robertson2004EPJAP} yielding $\kappa = 2.45$. In 3R-stacked MoS$_2$ bilayers, the electron is localized in one layer while the hole is shared between the two layers [Fig.~\ref{fig1}(c)]. In other words, the conduction-band minimum at K is layer-selective while the $\Gamma$ valence states are more strongly interlayer-hybridized. As a result, the average vertical electron-hole separation is approximately half the interlayer spacing ($\sim$0.62 nm),\cite{He2014PRB_Stacking} giving $d \sim 0.31$ nm and a dipole moment of $\sim$0.31 $e \cdot$nm.\cite{Liang2022PRX_Asymmetric} The resulting Boltzmann suppression factor ($U$) is plotted in Fig.~\ref{fig4}(b). In the rate-limited framework, annihilation is governed by short-range Auger-like coupling that requires significant wavefunction overlap,\cite{Erkensten2021PRB_DarkEEA,Erkensten2021PRB_Interaction} so the relevant interaction length scale is on the order of exciton size. Using a typical exciton Bohr radius of bilayer MoS$_2$ ($\sim$1 nm),\cite{Yu2015SciRep_Dielectric,Cheiwchanchamnangij2012PRB_QP} we obtain $V_{\mathrm{dd}}(r) = 52.7$ meV and $U = 0.13$, comparable in magnitude to the measured ratio $\gamma_{\mathrm{EEA},3\mathrm{R}}/\gamma_{\mathrm{EEA},2\mathrm{H}} \approx 0.35$. Notably, the measured ratio is reproduced by $r \approx 1.26$ nm, which is close to the theoretically predicted effective Bohr radius ($\sim$1.3 nm).\cite{Cheiwchanchamnangij2012PRB_QP} This agreement supports the interpretation of $r$ as an effective exciton-exciton interaction length scale. We model the dipolar repulsion using the electrostatic interaction between two out-of-plane dipoles in an effective dielectric environment, which provides an order-of-magnitude estimate for the short-range suppression factor. More refined treatments incorporating 2D screening (e.g., the Rytova--Keldysh potential)\cite{Cudazzo2011PRB_Screening} are expected to renormalize $V_{\mathrm{dd}}(r)$ but do not alter the central conclusion that dipolar repulsion reduces close-encounter probability and suppresses EEA. Overall, repulsive dipolar interactions generated by spontaneous polarization in 3R MoS$_2$ provide a quantitative, physically plausible mechanism for the additional suppression of EEA relative to 2H bilayers, highlighting the role of layer polarization in exciton-exciton annihilation.

\section{Conclusion}
Fluence-dependent ultrafast optical spectroscopy revealed stacking-dependent EEA rates and enabled a direct comparison of annihilation rates in CVD-grown MoS$_2$ monolayers and in 2H- and 3R-stacked MoS$_2$ bilayers. By analyzing the density-dependent A-exciton bleaching dynamics using a bimolecular recombination model, we extracted EEA rates of $(9.14 \pm 0.85)\,\times 10^{-2}$\,cm$^2$\,s$^{-1}$ for the monolayer, $(1.43 \pm 0.37)\,\times 10^{-2}$\,cm$^2$\,s$^{-1}$ for the 2H bilayer, and $(5.03 \pm 0.99)\,\times 10^{-3}$\,cm$^2$\,s$^{-1}$ for the 3R bilayer. The reduced annihilation rates in the bilayers are attributed to enhanced dielectric screening and the emergence of a phonon-assisted indirect annihilation pathway associated with the indirect band gap of bilayer TMDs. Beyond this layer-number effect, we observed an additional suppression of EEA in 3R MoS$_2$ bilayers arising from spontaneous polarization induced by inversion-symmetry breaking. The resulting dipolar excitons, formed through out-of-plane electron-hole separation, experience repulsive dipole-dipole interactions that reduce the probability of exciton-exciton encounters. Prior reports of superdiffusive exciton transport in 3R MoS$_2$ bilayers\cite{Agunbiade2025ACSNano_Superdiffusive} may appear inconsistent with the reduced EEA observed here when interpreted within the diffusion-limited picture.\cite{Yu2016PRB,Goodman2020JPCC,Mouri_PRB_nonlinearPL} However, EEA in two-dimensional semiconductors is not universally governed by diffusion alone;\cite{Yuan2015Nanoscale,Goodman2020JPCC} it can instead enter a rate-limited regime in which the bottleneck is the short-range Auger-like annihilation probability, determined by microscopic carrier overlap and interaction potentials. Within this framework, enhanced transport does not necessarily imply stronger EEA. Our observation of additional EEA suppression in the polar 3R stacking, together with the quantitative agreement of a dipolar-repulsion model, supports a rate-limited annihilation picture in 3R MoS$_2$ bilayers. The EEA rates obtained here should also be useful for modeling optoelectronic device operation, as they provide a quantitative parameter for this additional decay channel.\cite{Wu2022SciAdv_UltrafastPV} These findings indicate that stacking configuration and built-in polarization can be leveraged to mitigate exciton-exciton interactions and the associated nonlinear loss processes in two-dimensional semiconductors, thereby supporting higher sustainable exciton densities in 3R bilayers and offering potential advantages for bright emitters, nonlinear optics, and ferroelectric optoelectronic platforms.

\section{Methods}
\paragraph{Sample preparation.}
Bilayer MoS$_2$ samples were synthesized by atmospheric-pressure CVD. Si/SiO$_2$ substrates (285 nm oxide thickness) were placed face-down above an alumina crucible containing $\sim$2 mg of MoO$_2$ powder mixed with CsBr in a weight ratio of $\sim$1:0.04. This mixture was designated as the precursor. The crucible containing the MoO$_2$ precursor and substrates was loaded at the center of a quartz tube. Another crucible containing $\sim$50 mg of sulfur powder was placed upstream, where heating tape was wrapped around the tube. After flushing the tube with 500 standard cubic centimeters per minute (sccm) of ultrahigh-purity argon gas, the precursor was heated to 780 $^\circ$C (with a ramp rate of 40 $^\circ$C/min), and the sulfur was heated to $\sim$200 $^\circ$C by the heating tape under an 80 sccm argon gas flow. The reaction was carried out for 3 min. After growth, the heating tape was removed, and the furnace was opened to allow rapid fan-assisted cooling to room temperature.

\paragraph{Static optical spectroscopy.}
Raman measurements were conducted using a Jobin-Yvon T64000 triple spectrometer equipped with 1800 grooves/mm gratings and a liquid-nitrogen-cooled CCD detector (Symphony Horiba JY). All measurements were performed at room temperature using a 532 nm continuous-wave laser coupled into an upright microscope in a backscattering configuration. The 532 nm laser beam was focused through a 100$\times$ objective (NA = 0.9), yielding a beam spot of $\sim$1 $\mu$m with an incident laser power of $\sim$0.3 mW. Linear reflectance spectra ($R_0$) were measured at the same sample locations using a laser-driven broadband light source (Energetiq Inc.).

\paragraph{Ultrafast optical spectroscopy.}
Transient reflectance measurements were performed using a Ti:sapphire oscillator (Micra, Coherent) seeding a Ti:sapphire regenerative amplifier (Legend USP-HE, Coherent) operating at a 1 kHz repetition rate. The amplifier delivered $\sim$57 fs pulses centered at 800 nm with a pulse energy of 1.8 mJ. A broadband white-light continuum probe was generated by focusing a small portion of the 800 nm beam into a 2-mm-thick sapphire window. Pump pulses at 3.1 eV were produced via second-harmonic generation of the 800 nm fundamental. Collinear pump and probe beams were directed into a custom-built upright microscope equipped with a 36$\times$ reflective objective, resulting in pump and probe spot diameters of $\sim$5 $\mu$m at the sample. The pump fluence was adjusted using neutral-density filters. The reflected probe spectrum was dispersed by a spectrograph (Shamrock 303i, Andor) and detected with an EMCCD (Newton, Andor). The pump and probe pulses were cross-polarized, and a polarizer placed before the spectrometer was used to suppress pump scattering in the transient reflectance map. At each pump-probe delay, the transient reflectance was calculated as $\Delta R/R_0 = (R_{\mathrm{pump}} - R_0)/R_0$, where $R_{\mathrm{pump}}$ and $R_0$ denote the reflected probe intensities with the pump on and off, respectively.

The initial exciton density per layer, $n_0$, was estimated using an optical absorption of $\sim$25\% per MoS$_2$ layer at the pump wavelength.\cite{Canales2023ACSNano_StrongCoupling} Substrate interference was accounted for by the field-intensity reduction factor $I/I_0 = |1 + r_{\mathrm{eff}}|^2 \approx 0.1$ for 285-nm-thick SiO$_2$ on Si at the pump wavelength. Here,\cite{Li2012ACSNano_RamanThickness} $r_{\mathrm{eff}} = (r_{12} + r_{23}\Omega_2)/(1 + r_{12}\Omega_2)$ with $r_{ij} = (\tilde{n}_i - \tilde{n}_j)/(\tilde{n}_i + \tilde{n}_j)$, where subscripts 1, 2, and 3 denote air, SiO$_2$, and Si, respectively. The phase factor is given by $\Omega_2 = \exp(-4\pi i \tilde{n}_2 d_2/\lambda_{\mathrm{pump}})$, where $\tilde{n}_2$ and $d_2$ are the complex refractive index and thickness of the SiO$_2$ layer, respectively, and $\lambda_{\mathrm{pump}}$ is the pump wavelength. The complex refractive indices of SiO$_2$ and Si were taken from elsewhere.\cite{RodriguezDeMarcos2016OME,Aspnes1983PRB_OpticalConstants}

\begin{acknowledgement}

This work was supported by the Center for Nanophase Materials Sciences (CNMS), which is a US Department of Energy, Office of Science User Facility at Oak Ridge National Laboratory. Partial support for this work was provided by the Los Alamos National Laboratory Laboratory Directed Research and Development (LDRD) program (Project number: 20260340ER). Los Alamos National Laboratory, an affirmative action equal opportunity employer, is managed by Triad National Security, LLC for the US DOE NNSA, under contract no.~89233218CNA000001.

\end{acknowledgement}

\begin{suppinfo}

The Supporting Information is available free of charge at the publisher's website.

Raman spectroscopy of monolayer, 2H-, and 3R-stacked bilayer MoS$_2$ (Note~S1 and Fig.~S1); pump-probe signal and exciton-exciton annihilation rates in monolayer regions adjacent to 2H and 3R bilayers (Note~S2 and Fig.~S2); pump-fluence dependence of the pump-probe signal (Note~S3 and Fig.~S3); $F_{\mathrm{pump}}$-dependent fast and slow decay constants for monolayer, 2H, and 3R bilayer MoS$_2$ (Table~S1); exciton-exciton annihilation rates extracted at different $F_{\mathrm{pump}}$ (Table~S2).

\end{suppinfo}

\bibliography{Park_EEA}

\end{document}